\documentclass[10pt,preprint2]{emulateapj}
\usepackage{latexsym}

\newcommand{\mymail}{oysteol@astro.uio.no}
 
\begin{document}

\altaffiltext{1}{also at Center of Mathematics for Applications, University of 
Oslo, P.O. Box 1053 Blindern, N-0316 Oslo, Norway}

\title{The temperature diagnostic properties of the Mg I 457.1 nm line}

\author{{\O}ystein Langangen}
\author{Mats Carlsson\altaffilmark{1}}

\affil{Institute of Theoretical Astrophysics, University of Oslo, 
P.O. Box 1029 Blindern, N-0315 Oslo, Norway}
\email{\mymail}

\begin{abstract}
  We analyze the important formation processes for the 
  Mg~{\small{I}}~457.1~nm~line.
  This line is an intercombination line and
  the source function is
  close to the local thermodynamic equilibrium (LTE) value. 
  The strong coupling to the local temperature and the relatively high
  population of the lower level (the ground state of Mg I) makes this line 
  an ideal candidate for temperature diagnostics in the lower 
  chromosphere/temperature minimum region.
  Linking the temperature probed 
  to an absolute physical height is non trivial because of Non-LTE ionization. 
  We analyze the Non-LTE effects
  and find that photo-ionization from the lower energy levels 
  together with cascading collisional 
  recombination dominate the ionization balance. Taking properly into account
the line-blanketing in the UV is essential for obtaining the right photoionization rates.
  The identification of the main Non-LTE effects in the line 
  allows us to construct a ``quintessential'' model atom,
  ideal for computationally demanding tasks, e.g. full 3D  
  and/or time-dependent radiative transfer.
  Furthermore we analyze the diagnostic potential to temperature of this line
  in solar-like atmospheres, by synthesizing the line from a series of parametrized 
  atmospheric models. These models have been constructed with fixed 
  effective temperature, but with a variable heat term in the energy equation
  to obtain a chromospheric temperature rise at different heights.
  We conclude that the line has a significant potential in the diagnostics of
  the lower chromosphere temperature structure, especially for cooler atmospheres,
  such as sunspots.
\end{abstract}

\section{Introduction}
\label{sec:intro}
Deriving the solar atmospheric temperature structure from the emergent
spectrum is a classical problem \citep[e.g.][]{HSRA,val3c,spotm,FALC}.
It became clear quite early that the 1D semi-empirical models would need 
an increasing temperature in the chromosphere, to explain the increasing 
radiation temperature with increasing opacity in lines \citep{1961Athay}.
The existence of such a static temperature rise in the real sun
has, however, been questioned \citep{radyn2}. Hence, it is of great importance 
to have a good temperature diagnostic for the lower solar chromosphere.

One such candidate is the Mg {\small{I}} 457.1 nm~line
(from now on the 457.1 line). 
Note that we here use the common practice in astronomy to use
the wavelength in air for naming lines, in vacuum the
wavelength is 457.2~nm.
The 457.1 line arises from the forbidden transition
$2p^{6}\,\,3s^2\,\,^{1}\!S \leftrightarrow 2p^{6}\,\,3s\,3p\,\,^{3}\!P^{o}$,
hence the source function should be tightly coupled to the local
temperature \citep[e.g.][]{MgLTE,Mgheight,heasley,mauas}
due to the dominance of collisions in forbidden lines.
The fact that the source function is highly coupled to the local 
temperature makes the 457.1 line ideal as a diagnostic of the temperature 
at the height of formation of the line. Early attempts to use this line as
a temperature diagnostic include the work of \citet{white,Mgfirst,1974Altrock,
1977Rutten,heasley,mauas}. The most important result from this work is that the 
457.1 line is formed at around 500 km height in typical quiet sun
1D semi-empirical models, and hence the line will probe the temperature 
minimum region. Furthermore, the 457.1 line has also been used as a temperature 
diagnostic for cooler models such as sunspot models \citep{1987Lites}.
The modelling often gives a non-observed central reversal in the line 
core leading to the conclusion that the chromospheric temperature 
rise must be at a higher position. This result has, however, 
been questioned \citep{mauas} due to possible uncertainties in the 
atomic model. More comprehensive Mg {\small{I}} model atoms have been published 
like the 66-level atom of \citet{mg66p} and the 83 level atom of 
\citet{1998Zhao}. The 457.1 line has also been used to derive the 
temperature structure in solar features such as magnetic elements 
\citep{1995Briand} and flares \citep[e.g.][]{1990Metcalf1,1990Metcalf2}.

In this paper we present analysis and results of the temperature diagnostic 
properties of the 457.1 line. In \S\ref{sec:Form} we
analyze the important formation processes of the 457.1 line. As a result of
this work we present a minimal realistic model atom (a quintessential
model atom) for the 457.1 line in \S\ref{qma}. 
In \S\ref{sec:radyn} we parametrize the
model atmospheres by solving the radiation hydrodynamics equations
with a prescribed heating per unit volume. The atmospheres are allowed to 
relax and the resulting atmospheres are used to solve the Non-LTE
problem for the 457.1 line.
Finally we summarize our results in \S\ref{Con}.
\begin{figure*}[!ht]
\includegraphics[width=\textwidth]{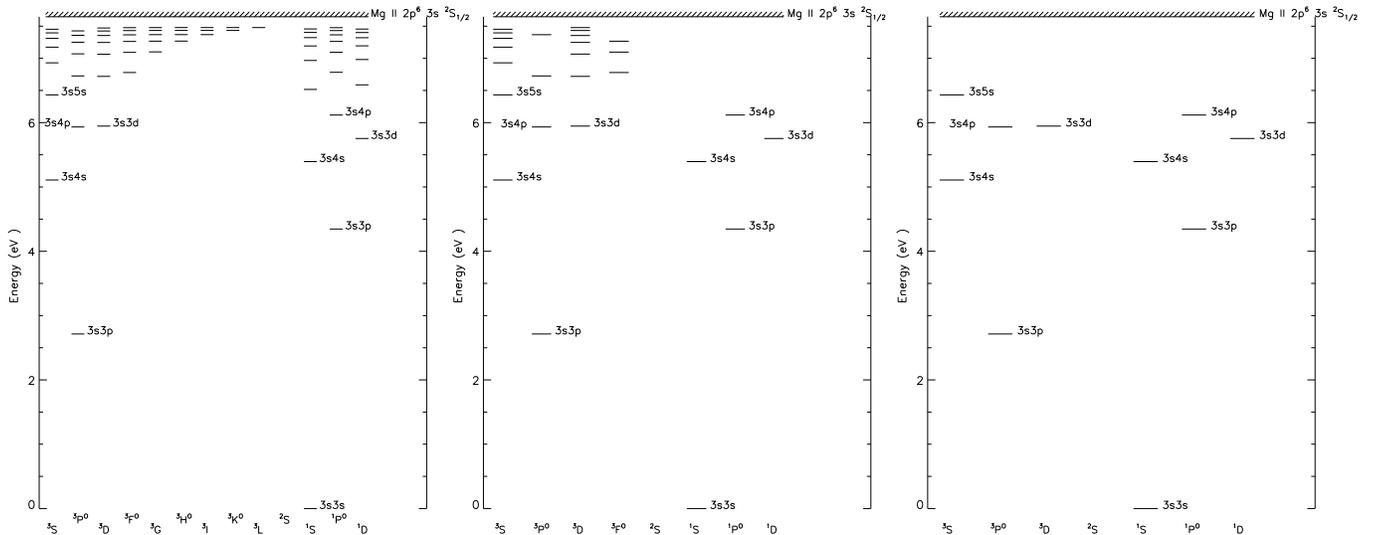}
\caption{Termdiagram of the 66 level atom ({\it left panel}), 
the 27 level quintessential atom ({\it center panel}), and the 
11 level test atom ({\it right panel}). To avoid confusion
we have refrained from labeling levels with energies
above 6.5 eV.}
\label{term}
\end{figure*}
\section{Formation processes for the 457.1  line}
\label{sec:Form}
To understand the important formation processes for the 457.1 line
we perform and analyse 1D radiative transfer calculations.
\subsection{Atmospheric models}
For this analysis we use two semi-empirical 1D models, 
the FAL~C model \citep{FALC} and the Spot~M model \citep{spotm}.
These models represent an average quiet sun atmosphere
and an average sunspot umbra atmosphere (for a sunspot with an umbral radius 
greater than 5\arcsec{} from the middle of the sunspot activity cycle), respectively.
At the formation height of the 457.1 line, magnesium is predominantly
neutral in the sunspot model and mostly ionized in the FAL~C model.
This large difference in the ionization makes these two models ideal
for testing the formation processes in different atmospheric conditions.
\begin{figure*}[!ht]
\includegraphics[width=\textwidth]{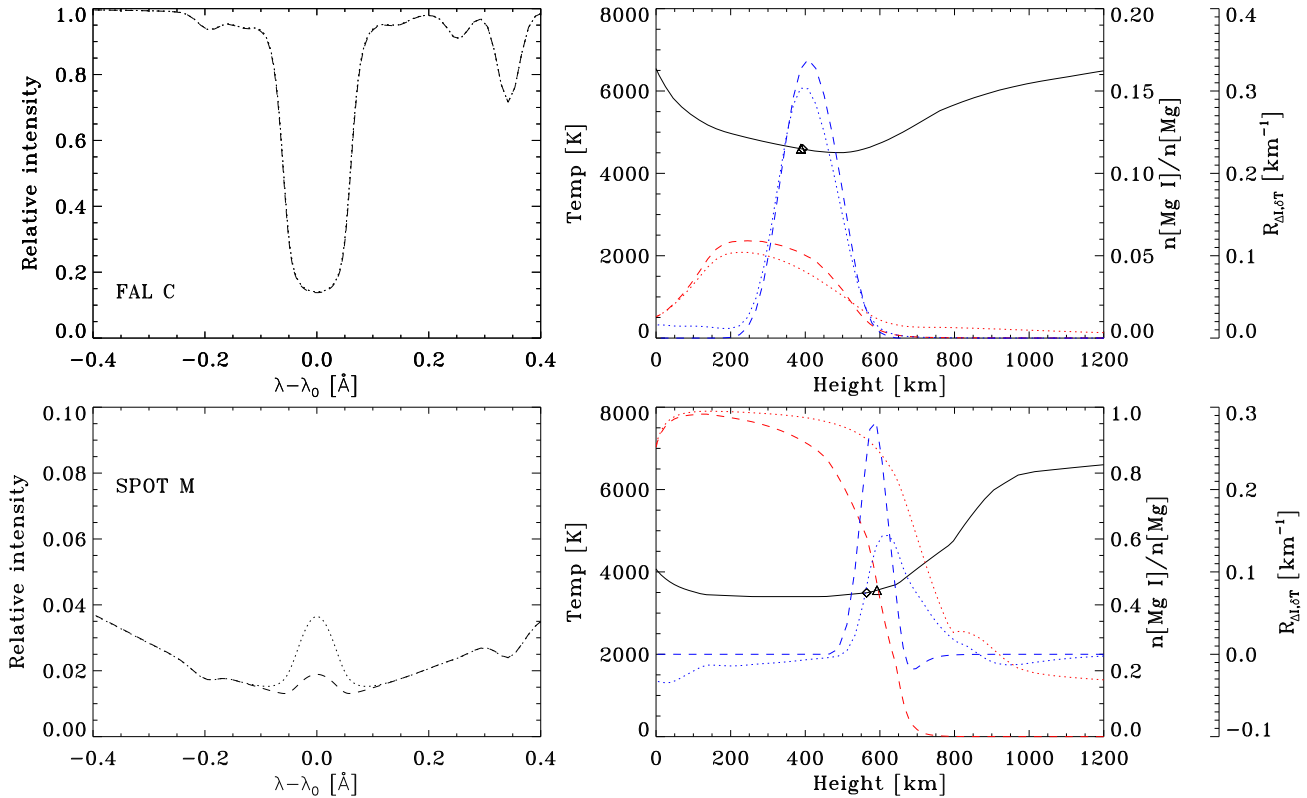}
\caption{FAL~C ({\it upper panels}) and Spot~M ({\it lower panels}). 
{\it Left panels}: Emergent profiles in LTE ({\it dashed}) and 
NLTE ({\it dotted}). {\it Right panels}: temperature ({\it black solid})
intensity response functions to temperature calculated in LTE ({\it blue
dashed}) and NLTE ({\it blue dotted}), and neutral fraction in
LTE({\it red dashed}) and NLTE ({\it red dotted}). The diamonds show where
$\tau_\nu$=1 in LTE and the triangles show
similar results in the NLTE case. Note the difference in the Y-axis for the
intensity, neutral fraction and response functions between the two
atmospheres. Relative intensity is relative to FAL~C continuum intensity 
in both cases.}
\label{prof}
\end{figure*}

\subsection{Model atom}
The atomic model used in this analysis was published by \citet{mg66p}.
In short, this is a 66-level atom, including 65 Mg I levels and one Mg II
level.The model atom include all levels up to $n=9$ and also the $n=10$
levels for the Mg I $2p^{6}\,\,3s\,10s\,\,^{3}\!S$,
Mg I $2p^{6}\,\,3s\,10s\,\,^{1}\!S$, and 
Mg I $2p^{6}\,\,3s\,10p\,\,^{1}\!P^{o}$, since these levels have so big
quantum defects that their energy is lower than the  $n=9$ energy levels with
high l values.
Furthermore, the model atom includes 65 bound-free transitions and 
315 bound-bound transitions. Only one intercombination transition is included,
$2p^{6}\,\,3s^2\,\,^{1}\!S \leftrightarrow 2p^{6}\,\,3s\,3p\,\,^{3}\!P^{o}$,
which in fact is the transition giving rise to the 457.1 line (note that the
data for this line is taken from \citet{mauas}). Also note that there is no
distinction between the singlet and the triplet systems for levels with
$l\ge 3$, i.e. similar levels are considered to be internally Boltzmann populated.
Finally, for some levels (most notably the upper level of the 457.1 line), 
the fine structure has been combined into
"terms", with the individual levels assumed to be Boltzmann populated with
respect to each other and the energy taken as the weighted mean energy
with the statistical weight of each of the fine structure 
levels as weight factor. A termdiagram of this model atom can be seen in Fig.\ref{term}.

In these calculations we adapt a Mg abundance of
7.53 on a logarithmic scale \citep{2005AGS}. All other heavy element
abundances are taken from the same source.
\subsection{Formation processes in the FAL~C atmosphere}
Using the Mg model atom and the FAL~C model atmosphere, we solve the equation
of radiative transfer and statistical equilibrium using the
MULTI code \citep{multi}. Background opacities from nearby lines are treated
with a scattering term included in the source function in a two-level
approximation. The line data for these lines were retrieved from the 
Vienna Atomic Line Database (VALD) \citep{1995Piskunov,vald,1999Ryabchikova}.
After solving the NLTE problem, we take into
account the fine structure of the upper level of the 457.1 line by 
distributing the term population over the sublevels solving the
transfer equation again for the 457.1 transition.

Since the bound--free edges of many of the important lower levels occur 
in the "UV line haze", it is necessary to account for the resulting line 
blocking when calculating the bound-free opacity. This is done by using 
an opacity sampling method which reads a precomputed table of line
opacity as function of wavelength, temperature 
and electron density \citep{2005Collet}.

\begin{figure}[!ht]
\includegraphics[width=0.5\textwidth]{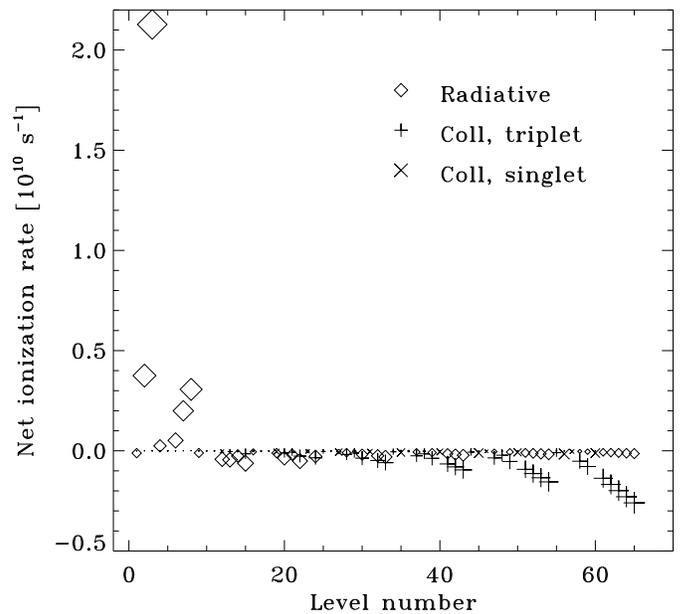}
\caption{Net ionization rates into Mg II from the different levels of Mg I 
at $\tau_\nu=1$ in the 457.1 line in
the FAL~C model. The symbol size is scaled to be proportional to the
rate. Radiative rates ({\it diamonds}), collisions
from singlet levels ({\it crosses}) and triplet levels ({\it pluses}).}
\label{rat}
\end{figure}

We test the NLTE effects in the line formation processes,
by comparing the LTE solution with the NLTE solution. The emergent
profiles in the two cases can be seen in the upper left panel of Fig.\ref{prof}.
The difference in the two line profiles is very small. The 
response-function to intensity given a change in temperature 
\citep[from now on denoted $R_{\Delta I,\delta T}$, see][for details]{1986Magain,
2005Astrid}, is seen in 
the upper right panel in Fig.\ref{prof}. There is not much difference
between the LTE and the NLTE $R_{\Delta I,\delta T}$, but this small difference
is actually a coincidence.
Since the lower level in the 457.1 line is the ground state of Mg I,
the population density of this line is set by the ionization
fraction of the atom. The  fraction of Mg I atoms to the total number of Mg 
atoms is seen in the upper right panel of Fig.\ref{prof}. It is clear that 
the relative ionization in LTE compared to NLTE deviate 20\% or more, 
but it turns out that the under--ionization at greater atmospheric 
heights to some extent cancel the over--ionization at lower heights in the 
atmosphere. We investigate these NLTE effect by comparing the rates 
into the Mg II level from each Mg I level, 
see Fig.\ref{rat}. As indicated by Fig.\ref{rat}, the dominating effects in
the ionization balance of magnesium is radiative ionization from the lower 
levels, (which dominate because of their relatively high population) 
especially the Mg I $2p^{6}\,\,3s\,3p\,\,^{1}\!P^{o}$ level (level number 3), 
which has significantly larger photo-ionization crossection compared to other 
similar levels. Furthermore, the high energy triplet energy levels play
an important role in the ionization balance through collisional recombination
\citep{mg66p,2000Shimanskaya}.
\subsubsection{Testing the sensitivity of the ionization fraction}
To further test the dominating processes in the ionization balance of magnesium,
we perform a series of test runs with different degree of 
realism in the radiative transfer. 
We test the two main NLTE effects, namely the photoionization
and the collisional recombination. For the first effect we
run our model atoms with and without taking into account the 
``UV line haze''. For the second effect, we construct a model atom which 
does not contain the recombination ladder, essential for the cascading 
recombination. This is done simply by removing 
all energy levels, except the Mg II level, with energy above 6.5 eV, 
see Fig.\ref{term} right panel for details on the resulting 11 level atom.
The 11 level model atom is quite similar to the older models of the
Mg {\small{I}} atom. 
\begin{figure}[!ht]
\includegraphics[width=0.5\textwidth]{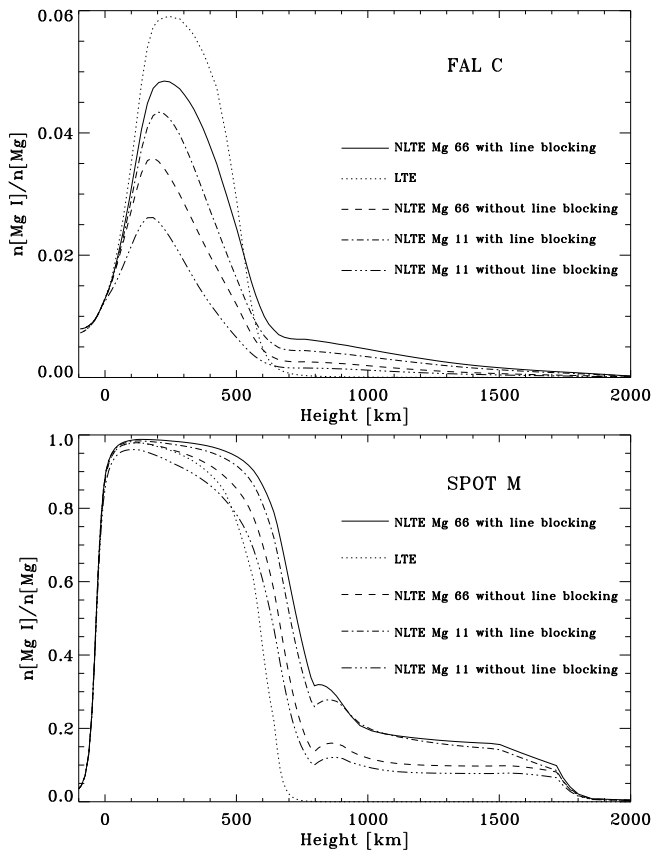}
\caption{neutral fraction in the FAL~C ({\it upper panel}) 
and Spot~M ({\it lower panel}) models
with different level of detail included in the radiative 
transfer calculations. Note the difference in scaling on the Y-axis between
the two panels. Line styles are defined in the figure.}
\label{ion}
\end{figure}
The results of these tests for the FAL~C and the Spot~M atmospheres 
are seen in Fig.\ref{ion} and can be summarized as 
follows: The largest effect comes from neglecting line blocking with
a similar, but slightly smaller, change in the results if
the cascading recombination is neglected.
It is clear that the NLTE
effects are significant, and it is very important that they are included
in a realistic manner. If the NLTE effects are included, but with
a lacking realism, the deviations from LTE are generally amplified.
Finally, to understand the under-ionization high in the atmosphere,
and the over-ionization in the lower atmosphere we turn our attention 
to the radiation field. If $J_\nu$ is bigger than 
$S_\nu$ we have more photons absorbed than emitted and vice versa.
Investigating the radiation field for the most important bound--free
transition, we see that $J_\nu$ is larger than
$S_\nu$ where we have over-ionization and smaller than $S_\nu$ higher
in the atmosphere where we have under-ionization.

\subsection{Formation processes in the Spot~M atmosphere}
\begin{figure}[!ht]
\includegraphics[width=0.5\textwidth]{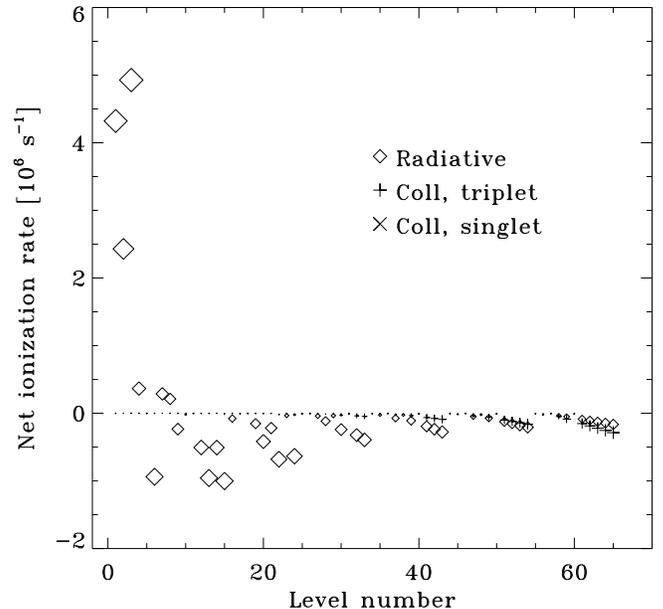}
\caption{Net ionization rates into Mg II from the different levels of Mg I at
$\tau_\nu=1$ in the 457.1 line in
the Spot~M model atmosphere. The symbols are scaled to be proportional to the
rate. Radiative rates ({\it diamonds}), collisions
from singlet levels ({\it crosses}) and triplet levels ({\it pluses}) are plotted.}
\label{rat2}
\end{figure}
\begin{figure}[!ht]
\includegraphics[width=0.5\textwidth]{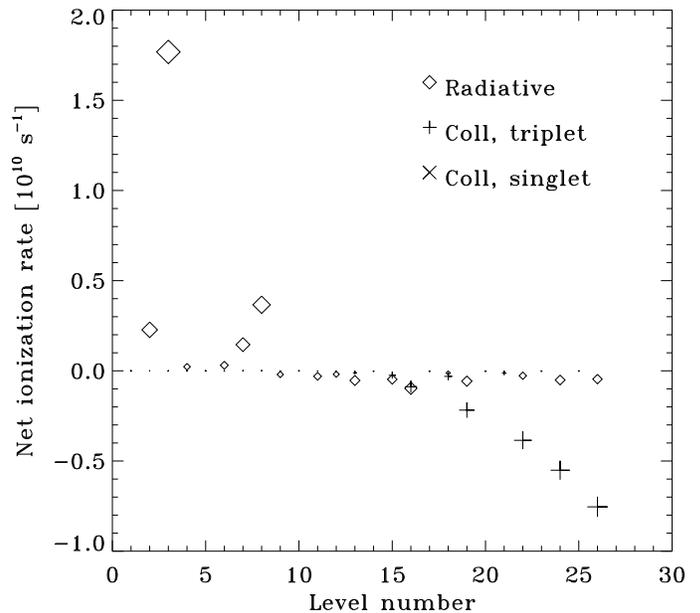}
\caption{Net ionization rates into Mg II from the different levels of Mg I at
$\tau_\nu=1$ in the 457.1 line using the quintessential 27 level mg atom in
the FAL C model atmosphere. The symbols are scaled to be proportional to the 
rate. Radiative rates ({\it diamonds}), collisions
from singlet levels ({\it crosses}) and triplet levels ({\it pluses}) are plotted. Compare to Fig.\ref{rat}.}
\label{rat3}
\end{figure}
\begin{figure*}[!ht]
\includegraphics[width=\textwidth]{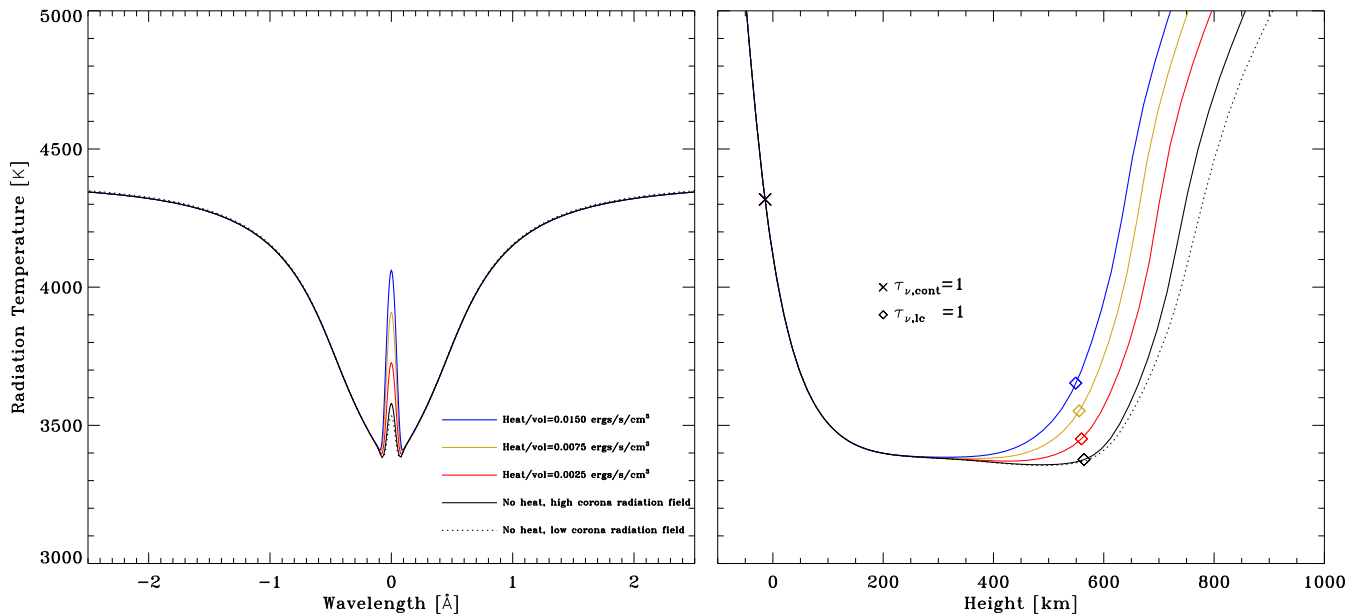}
\caption{Emergent profiles for the different model atmospheres are plotted
({\it left panel}). The temperature structure of these models are shown 
({\it right panel}). Line styles are showing the different atmospheric models
and the corresponding emergent profile. The atmospheres are denoted by the
logarithm of the column mass (in cgs units, g cm$^{-2}$) of the chromospheric temperature increase.}
\label{spot_vol}
\end{figure*}
Similar to the FAL~C model we solve the equations of radiative transfer 
for the Spot~M model. The resulting line profiles, LTE and NLTE, 
are seen in the lower left panel of Fig.\ref{prof}. The difference between the two 
profiles is significant, especially compared to the similar FAL~C profiles.
The NLTE solution shows a pronounced increase in the central reversal 
in the line center of the line. This is basically caused by the higher
geometric position of the monochromatic optical depth in NLTE, where
the temperature is higher compared to the temperature at the height of
formation in LTE. The $R_{\Delta I,\delta T}$ is much wider in NLTE compared 
to the LTE case. Furthermore, the $R_{\Delta I,\delta T}$ has contributions 
in NLTE from high regions
where the temperature has a significant increase. The reason for the 
increasing formation height in the Spot~M model is an increase in the 
fraction of neutral Mg, see Fig.\ref{prof} lower right panel.
To understand this deviation in ionization fraction, we plot the 
net rates into the Mg II levels, see Fig.\ref{rat2}.
As for the FAL~C model, we get big contributions from the lower levels 
of the atom, but also the higher excited levels contribute significantly
to the ionization fraction, but with a negative contribution.
This, together with the cascading collisional recombination, gives
much lower ionization fraction for the Mg I atom in the Spot~M model.
From Fig.\ref{ion} lower panel, one can see that the ionization fraction is
less sensitive to the number of levels in the atom, but much more
so for the treatment of line blocking. This is due to the reduced relative 
importance of the cascading recombination, compared to radiative transitions,
see Fig.\ref{rat2}

\section{Quintessential model atom}
\label{qma}

From the above analysis in \S\ref{sec:Form} we have identified the
dominating processes in the formation of the 457.1 line.
It is then a reasonable question if it is possible to 
construct a ``quintessential'' model atom (the smallest atom which reproduces
the important physical processes) for this line. 
Since it is very important to get
the ionization fraction correctly for this line, such a quintessential atom
must contain a recombination ladder, which satisfactorily mimics the cascading
recombination seen in the 66 level model atom. 

Following the general 
procedures outlined by \citet{2008Bard}, we construct a quintessential 
457.1 line model atom. First, we remove all singlet levels with energy 
levels above 6.5 eV. This is justified by the small contribution to
the overall ionization fraction from these levels, as seen in 
Fig.\ref{rat} and in Fig.\ref{rat2}.
If two levels have a difference in effective quantum number
${\Delta}q\leq 0.2$ and at the same time are almost Boltzmann populated 
(${\Delta}\beta \leq 0.1$) in both in the FAL~C model and in the Spot~M model
they are merged. The new model atom is tested in the two atmospheric models,
and only accepted if the maximum relative deviation in the emergent intensity 
(compared to the full atom) is less than 25\%. This procedure is 
repeated until all levels have been tested. The resulting model atom
is reduced to a 27 level atom; see Fig.\ref{term}, which is a reduction
to less than half the size of the original atom. 

Further reduction is
possible, but only with more liberal acceptance conditions, which
would affect the realism of the model atom. The new model atom mimic the
recombination ladder in the full model atom, by super levels as seen in
Fig.\ref{rat3} and Fig.\ref{rat}.

The quintessential model atom was also tested against the original atom 
using a series of dynamic atmospheres taken from \citet{radyn3}.
This test gives a deviation between the lineprofiles of less than 3\% between
the two models atoms, and deviations in the bisector Doppler shifts 
of usually less than 10 m\,s${}^{-1}$ and always less than 30 m\,s${}^{-1}$.

\section{Parametrization of the temperature diagnostics}
\label{sec:radyn}

In the above analysis we have seen that the 457.1 line shows a strong central 
reversal in a cool sunspot atmosphere, but in a hotter quiet sun 
model there is no such reversal. The reason for the lack of reversal in the hotter model is the
high ionization fraction in such models, which deplete the number densities
of the lower level in the 457.1 line and put the formation height below the 
chromospheric temperature rise. From here on we focus on cool 
sunspot models, since the 457.1 line has a clear diagnostic potential for 
such models. We are especially interested in the possible bound the central 
reversal put on the position of the temperature rise in such models.

To investigate the diagnostic properties of the line further,
we construct a series of 1D self consistent radiative hydrodynamics 
model atmospheres. We use the radyn code \citep{radyn1,radyn2,radyn3,radyn4}, 
to construct these model atmospheres. 
In short this code solves the mass, momentum, 
and energy conservation equations together with the detailed equations
of radiative transfer for three of the important elements in the energy balance
(H, Ca, and He) on an adaptive grid \citep{dorfi}. The boundary condition 
in the deep photospheric part (colum mass $\ge$ 15 g\,cm${}^{-2}$) of 
the atmosphere is taken from a 4000 K effective temperature model without 
convection generated by the MARCS code \citep{MARCS}. The temperature 
structure in these deep parts of the atmosphere is fixed by adding a new 
term to energy equation in this region. 

The incoming UV radiation field from the corona has to be specified at the 
upper boundary. To facilitate such a boundary condition we use the 
data of \citet{1991Tobiska} and follow the method used 
by \citet{1994Wahlstrom}. Since the model used by 
\citet{1994Wahlstrom} was based on a quiet sun corona, we expect that 
this model underestimate the coronal radiation field.
We have performed tests (a more thorough analysis of this effect will be given
in an upcoming paper) with different scaling of the incoming radiation
field, and found that the radiation field has profound effects on 
the chromospheric temperature structure. From these tests we conclude that 
the incoming radiation field should be scaled up by at least a factor 5.
With the above described boundary conditions, we allow the
atmospheres to converge to an equilibrium state, and the 
pressure at the top of the simulations is changed until there are no more
outflowing material. The resulting temperature structures for an active and
a quite corona are seen in Fig.\ref{spot_vol} right panel.
The difference in the chromospheric temperature structure due to scaling 
of the coronal radiation field is quite big, however the effect on the 
457.1 line is rather small due to the low formation height, 
see Fig.\ref{spot_vol} left panel. For the construction of the atmospheres used 
in this analysis we adopt a scaling factor of 5 for the coronal radiation field. 

We parametrize the position of the chromospheric temperature increase in the 
theoretical model atmosphere by adding a constant heat term in the energy 
equation. 
To avoid unrealistic temperature structures in the lower part of the atmosphere,
the heat term has to be defined per volume and not per mass. The latter 
will lead to an increase in temperature at very low atmospheric heights
\citep{2004Bard}. Since the heating term per volume will lead to a very high 
energy input in the corona, we allow a smooth transition to heat per mass above the
transition region in the original atmosphere. The exact position of
this smooth transition is changing the temperature structure in the
corona, and to some extent the upper chromosphere, but it has
no effect on the emergent 457.1 line profile. Furthermore, the heat term
leads to an expanding corona 
but flows due to this expansion at the formation height of the 457.1 line
are very small, typically less than 10 m\,s${}^{-1}$ and hence negligible.

We construct different atmospheres by adding different heat terms 
per volume to the energy equation as described above. 
In this way we construct a grid of
self consistent atmospheres with different position of the chromospheric 
temperature rise. Note that we use this parametrization of the 
temperature because it is only dependent on one parameter 
(the magnitude of the constant heat term added to the energy equation).
Furthermore, these models should be considered a parametrization of the 
position of the chromospheric temperature increase, and {\it not} a 
parametrization to evaluate the energy balance in the chromosphere. 

Some of the resulting temperature structures are seen in Fig.\ref{spot_vol}
right panel.
As seen in the left panel of Fig.\ref{spot_vol} the emergent 457.1 line
profile shows a significant increase in the central reversal with decreasing
height of the chromospheric temperature rise. Note that these line profiles
have been calculated using the quintessential modelatom and without including
the nearby lines (to avoid confusion).
The size of the reversal is increased 
by the increasing temperature at the monochromatic optical depth of the 
line center of the 457.1 line. Note that the position of the monochromatic 
optical depth is only slightly changed with the changes in the position of
the chromospheric temperature rise. 
The reason for this rather constant formation height is the high number
densities of the Mg I ground state. If the temperature is increased, the
relative change in the ionization fraction is changed only slightly due
to the high density of Mg {\small{I}} atoms. 

Note the inconsistency between
the emergent profile and the Eddington-Barbier approximation applied
on the temperature structures in Fig.\ref{spot_vol}. The reason for the
lacking correspondence is the deviation of the source function/temperature 
from a linear function in $\tau_\nu$. 
Since the Eddington-Barbier approximation is not very good,
it is not very instructive to invert the temperature profile from
such assumptions. However, the reversals are strong enough to 
discriminate between different models. We recommend a forward modeling
approach to the temperature diagnostics, similar to the work by 
\citet{1987Lites}. From these theoretical models 
the 457.1 line can be used to set lower bounds on the position of the 
chromospheric temperature increase. The sensitivity of such a discrimination
will of course depend on the sensitivity of the observations used for
the comparison.

\section{Conclusion}
\label{Con}
We have analysed the dominant NLTE effects of the 457.1 line in two solar
models (FAL C and Spot M), and
concluded that the ionization from the lower levels, especially the
Mg I $2p^{6}\,\,3s\,3p\,\,^{1}\!P^{o}$, is the main contribution to the
ionization. Furthermore, cascading collisional recombination in
the triplet system plays an important role in the NLTE formation 
processes in the 457.1 line. With the understanding of the formation
processes for the 457.1 line, it is possible to 
construct a ``quintessential'' model atom. We have presented a 27 level 
quintessential model atom (an atom with the minimum number of levels, 
that maintain the important physical processes). 

We have found that increasing the realism in the model atom does not
remove the non-observed central reversal in the Spot M model, as
claimed by \citet{mauas}.

Finally we have presented a series of sunspot models where the
height of the chromospheric temperature rise has been varied. These
models show that the height of formation of the 457.1 line is not
very sensitive to the height of the chromospheric temperature rise
but the central reversal is. The amount of emission in the line
core is thus a sensitive diagnostic of the lower chromosphere
temperature structure and a lack of central reversal puts a
strong constraint on the lowest possible position of the
chromospheric temperature increase.

\emph{Acknowledgments.} This research was supported by the Research Council of Norway through grant 170935/V30 and a grant of computing time from the Program for Supercomputing.
This research has made use of NASA's Astrophysics Data System.
\bibliographystyle{apj}
\bibliography{paper5}

\end{document}